\documentstyle[12pt,amssymb]{article}
\newcommand{\smdag}{\mbox {\tiny \dag}}

\begin{document}

\begin{center}
{\Large {\bf
A $(p/E)$ CALCULATION OF STRONG PIONIC DECAYS OF BARYONS
}}
\end{center}

\vspace{1cm}

\begin{center}
F. Cano$^{1}$, P. Gonz\'alez$^{1}$, B. Desplanques$^{2}$ and S. Noguera$^{1}$
\end{center}

\begin{center}
$^{1}$ Departamento de F\'{\i}sica Te\'orica and IFIC,\\
Centro Mixto Universidad de Valencia-CSIC\\
46100 Burjassot (Valencia), Spain\\ 
\vspace{1cm}
$^{2}$Institut des Sciences Nucl\'eaires, F-38026\\
Grenoble Cedex, France
\end{center}

\vspace{3cm}

\begin{abstract}
{\small Strong pionic decays of baryons are studied in a non--relativistic 
quark model framework via a convergent $(p/E)$ expansion of the transition 
operator. Results are compared to the ones obtained within a more
conventional $(p/m)$ expansion.}
\end{abstract}

\vfill

\vfill

\newpage

\section{Introduction}

      There is a current experimental and theoretical interest 
in the study of the baryon decays (see for instance \cite{BARYON95}).
Measurements which are planned at different experimental facilities, 
TJNAF (CEBAF), MAMI, ELSA, GRAAL, aim to answer open 
questions about the
baryon resonances, in particular having to do  with 
the position of the Roper resonance in the baryon spectrum and with 
the absence of strong experimental evidence for many resonances 
around 1.8-2 GeV. Both subjects are
closely related to the study of the baryon decay amplitudes. Some authors
have proposed that small couplings in the $N\pi$ production channel can  
provide an explanation for the missing states \cite{CAPSTICK93}. On
the other hand the analysis of the corresponding decays is essential
to any attempt to establish the true nature of the Roper resonance for
which several alternatives have been proposed (hybrid three quark+one
gluon state, first radial nucleon excitation...)
Hence, the study of
the $N\pi$ decay amplitudes becomes of great interest, as a test for the
proposed models and in order to have a guide-line for future experiments.
In this article 
we shall center our attention at the lowest energy decays for reasons
that will become clear later on.

	In a previous paper \cite{CANO95} we put the emphasis 
in the construction of the effective
transition operator for $B \rightarrow B' \pi$ decays, 
making clear not only the very relevant known 
role played by
the pion structure (otherwise said, by the $qqq \; q\bar{q}$
effective component of the baryon wave function) but also the need of taking
into account relativistic effects. Here we go further in this analysis
and show that 'better accounted' relativistic corrections through a convergent
$(p/E)$ (instead of $(p/m)$) expansion does not spoil  our results but
instead confirm in a more sound way our previous conclusions. 

	For the sake of completeness we briefly review next
the quark models used in ref. \cite{CANO95} to calculate the 
baryon wave functions which we
shall employ for estimating baryon decay transition amplitudes. There
are many other competitive models available in the literature. However
for our purpose here, the comparison of a $(p/E)$ expansion
calculation against a $(p/m)$ one, we do not feel our particular
choice to suppose any serious limitation.

\section{The Quark Models.}

	A precise description of the octet and decuplet baryonic
spectrum (with and without strangeness) including the second
excitation energy have been consistently 
obtained by solving the Schr\"odinger
equation with a potential containing apart from the 'minimal
ingredients' (confinement + (coulomb + spin--spin) one gluon exchange
interactions) 
a three quark phenomenological force \cite{DESPLANQUES92}.
Specifically the expression used for the potential is:

\begin{equation}
V = V^{(2)} + V^{(3)}
\end{equation}

\begin{eqnarray}
V^{(2)} & = & \sum_{i<j} \frac{1}{2} \left[ \frac{r_{ij}}{a^2} - 
\frac{\kappa}{r_{ij}} + \frac{\kappa}{6 m_{i} m_{j}} 
\frac{\exp (-r_{ij}/r_{0})}{r_{0}^{2} r_{ij}} \vec{\sigma}_{i} 
\vec{\sigma}_{j}-D
\right] \\ 
V^{(3)} & = & \sum_{i\neq j \neq k \neq i} \frac{1}{2}
\frac{V_{0}}{m_{i} m_{j} m_{k}} \frac{e^{-m_{0} r_{ij}}}{m_{0} r_{ij}}
\frac{e^{-m_{0} r_{ik}}}{m_{0} r_{ik}}
\end{eqnarray}

\noindent
where a, $\kappa$, $r_{0}$, $V_{0}$, $m_{0}$  are free parameters
fixed from the spectrum and the the quark masses are chosen to get the
baryon magnetic moments (see table 1). 
$r_{ij}$ is the interquark distance and the
$\sigma$'s are the Pauli matrices. $D$ is a constant to fix the
absolute value of the nucleon mass to its experimental value.

	Such a model
is adequate to solve altogether some endemic problems
concerning the baryon spectrum, say, a unified description of the
positive and negative parity states, the correct position of the Roper
resonances (first radial excitations) and the appearance of extra $\Delta$
negative parity states at a relatively low energy ($\Delta(1/2^+)$ at 1900 
MeV and $\Delta(5/2^+)$ at 1930  MeV). Moreover, when combined  with 
an improved transition operator,
the $B \rightarrow B' \pi$ decay widths for the lowest energy
non-strange resonances are in very good  agreement (to the
standard in the field) with the experimental data
\cite{CANO95}.

	As a side effect, the splitting between the two lowest
$\frac{1}{2}^{-}$ and $\frac{3}{2}^{-}$ states, is rather small for
all the baryons studied (related to the absence of one gluon exchange
tensor hyperfine or spin--orbit forces) 
and an unobserved proliferation of states
would come out if we were to push the model further up the second
excitation energy (maybe denoting the energy limit of the model
description or maybe indicating the presence of missing states).

	We shall compare the results from the ($V^{(2)}  +
V^{(3)}$) potential model with the ones obtained with a two-body
$V^{(2)}$ potential model with parameters fitted to get an
overall fit to the meson and baryon spectrum \cite{GIGNOUX85}, the aim
being to try to extract general features associated to 
'two--body' models and to make clear, if so, some bias of the results from our
'three--body' approach. The small core size typical in spectroscopic
models ($\langle r^2 \rangle^{1/2} \approx 0.47$ fm for $V^{(2)}$)
makes the speed of the quarks be
close to 1. As a matter of fact $(p/m) \gtrsim 1$.
In this sense a $(p/E)$ expansion
is more founded and incorporates automatically some relativistic
corrections.

        Our results are expressed in terms of the width for each process.
In all cases, the amplitudes have been calculated with the model wave
functions, but we have used the physical masses for the kinematical factors.
This is important for the $V^{(2)}$ potential model due to the
discrepancies of its predictions with the experimental masses.

	Several mechanisms have been proposed to study the 
$B \rightarrow B' \pi$ decays.

\section{Decay Mechanisms and Results.}

	The {\bf elementary emission model} considers the point-like pion
emission by one of the quarks of the baryon. The transition
operator is obtained via the non--relativistic reduction of the
$qq\pi$ interaction for which a  pseudovector form is
used. The $(p/m)$ order results have been published elsewhere \cite{CANO95}.
The predicted widths of some low-lying non--strange baryons 
in the elementary emission model are
summarised in tables 2a and 2b,
showing a general disagreement with experiment. Furthermore, 
the calculation of the quark matrix element 
$\bar{u}(p)\gamma_{\mu} \gamma_5 u(p)$
at the $(p/m)^2$ order has been worked out \cite{CANO94} (we do not use the 
recalculated value of $f_{qq\pi}$ in order not to have an even bigger 
contribution),
making clear the relevance
of the corrections and the lack of convergence hence raising serious 
doubts about the adequacy to proceed to such an expansion. 

	These problems can be partially palliated by proceeding to a $(p/E)$
expansion ($E=\sqrt{p^2 + m^2}$) of the transition
operator (for obvious reasons we shall work in momentum space). It
is very instructive to do it order by order 
so that we can check the convergence of
the ($p/E$) expansion and compare it to the ($p/m$) one.
Thus the amplitude at $(p/E)$ order, can be written as:
    
\begin{equation}
\langle B' (J_{z}'=\lambda), \pi^{\alpha} | H | B (J_{z}=\lambda)
\rangle = \frac{1}{(2 \pi)^{3/2}} \delta (\vec{P}_{B} - \vec{P}_{B'} -
\vec{k}) A_{\lambda}
\end{equation}

\begin{eqnarray}
\label{amplitudgordape}
A_{\lambda} & = & - \frac{3 i}{(2 \omega_{\pi})^{1/2}} 
\frac{f_{qq\pi}}{m_{\pi}} 
\int d\vec{p}_{\xi_{1}} \;  d\vec{p}_{\xi_{2}} \; \\
& &  
\left[ \Psi_{B'}(\vec{p}_{\xi_{1}},\vec{p}_{\xi_{2}} +
\sqrt{\frac{2}{3}} \vec{k}) \Phi_{B'} 
(M',M_{S}';I',M_{I}') \right]_{J'\lambda}^{*} \nonumber \\ 
&  &\;\;\;\;\;\;\;(\tau^{\alpha\; (3)})^{\smdag} 
\vec{\sigma}^{(3)} \left[
\vec{k} \left( 1 + \frac{\omega_{\pi}}{6 E'}\right)  \nonumber \right. \\
& &  \left. \;\;\;\;\;\;\;\;\;\;\;\;\;\;\;\;\;\;\;\;\;\; +
\frac{\omega_{\pi}}{2}
\sqrt{\frac{2}{3}} \left( \frac{\vec{p}_{\xi_{2}}}{E} +  
 \frac{\vec{p}_{\xi_{2}} + \sqrt{\frac{2}{3}} \vec{k}}{E'} \right)
\right] \cdot \nonumber \\
&  &   \;\;\;\;\;\;\; \left[ 
\Psi_{B}(\vec{p}_{\xi_{1}},\vec{p}_{\xi_{2}}) \Phi_{B} 
(S,M_{S};I,M_{I}) \right]_{J\lambda} 
\end{eqnarray}

where $J$,$J'$ are the total baryon angular momenta, 
$\vec{P}_{B,B'}$ and  $\vec{k}$ the three-momenta of the baryons
and pion respectively, $\omega_{\pi}$ ($m_{\pi}$) is the pion energy (mass), 
E (E') refers to the energy of the incoming (outgoing) quark in the emission,
$\Psi$
($\Phi$) stands for the momentum space (spin--isospin) 
wave function,
$\vec{\tau}^{(3)}$ ($\vec{\sigma}^{(3)}$) is the isospin (spin) Pauli
matrix acting on the emitting quark, and
$f_{qq\pi}$  is the coupling constant 
in the $qq\pi$ vertex and $\vec{p}_{\xi_{1}}$, 
$\vec{p}_{\xi_{2}}$ are the momenta associated to 
quark Jacobi coordinates.

	The first term in the intermediate 
square bracket appearing in (\ref{amplitudgordape}), 
arises (except for the $\frac{\omega_{\pi}}{6 E'}$ contribution)
from the spatial components
while  all the other terms, proportional to
$\omega_{\pi}$, arise from the time
component of the current which is coupled to the derivative of the pion
field. 
It is easy to verify that eq. (\ref{amplitudgordape}) has
well defined properties (change in sign) when exchanging the role of
the final and the
initial baryon. Let's also note that the pseudo-scalar $\pi q q $ coupling
leads to an expression similar to (\ref{amplitudgordape}) 
where $\omega_{\pi}$ is replaced
by the difference of the kinetic energies of quarks in the initial and final
states, which is nothing else than the corresponding contribution to 
$\omega_{\pi}$.

	Beyond the $p/E$ order, typical terms like 
$\vec{p}/(E+m)$, that appear in the Dirac spinors will be expanded as

\begin{equation}
\frac{\vec{p}}{E+m} = \frac{\vec{p}}{2E} ( 1 + \frac{p^2}{4 E^2} + ...) ,
\end{equation}

\noindent
whereas a $p/m$ expansion would give

\begin{equation}
\frac{\vec{p}}{E+m} = \frac{\vec{p}}{2m} ( 1 - \frac{p^2}{4 m^2} + ...) ,
\end{equation}

	As usual, we shall work in the center of mass system of the
decaying baryon, i.e. $\vec{P}_{B}=0$. The only free
parameter $f_{qq\pi}$ is fixed 
from the $NN\pi$ form factor at zero momentum
transfer ($|\vec{k}|=0$). Additionally, the $NN\pi$ form factor is predicted
at $|\vec{k}| \neq 0$ (both models do very well for low $|\vec{k}|$).
$f_{qq\pi}$ is determined to the order of the
calculation from $f_{NN\pi}(0)$.

	The results for the same set of 
widths are shown in tables 2a and 2b. By comparing columns four,
five and six (this one corresponding to the complete calculation for the 
pseudovector form) with column two and three, the better 
convergence of the $(p/E)$
calculation becomes obvious for both quark models. 
In the $p/E$ expansion, the $(p/E)^2$ 
correction make the result to tend smoother
to the final one, while in the $p/m$ expansion the $(p/m)^2$ correction 
either go in the wrong direction or make the result to overshoot the 
final one. Strictly speaking only for some cases: $\Delta(1232)
\rightarrow N\pi$, $N(1520) \rightarrow N\pi$, $N(1535) \rightarrow
\Delta \pi$, $N(1650) \rightarrow \Delta \pi$ and $N(1700) \rightarrow
N \pi$, a first order $(p/E)$ treatment is justified whereas for $N(1440)
\rightarrow \Delta \pi$ and $N(1520) \rightarrow \Delta \pi$ it should
be taken with caution. 
The situation becomes
more dramatic when considering higher excitations for which we do not
find a justification neither to a $(p/m)$  nor to a $(p/E)$ expansion. 
Let us note that for $N(1440)
\rightarrow N\pi$, $\Delta(1600) \rightarrow \Delta\pi$ decay amplitudes, 
the first
order is suppressed by the orthogonality of the baryonic radial wave
functions. 
The corrections $(p/m)^2$ obviously break the orthogonality argument,
hence the effect is large for these transitions. Instead, for corrections 
$(p/E)^2$, which are close to the unit operator, the orthogonality argument
seems to work again, hence the smaller effects.

	Differences with
the data require to look back at the reaction mechanism; in particular
one can wonder whether the pion structure may play or not a
significant role.

	The {\boldmath $^{3} P_{0}$} {\bf quark pair creation model}
considers the
creation of a $q\bar{q}$  pair in the hadronic medium that by later
recombination gives rise to the outgoing pion. The effective
transition operator is \cite{LEYAOUANC73,ROBERTS92}:

\begin{eqnarray}
T & = & -\sum_{i,j} \int \; d\vec{p}_{q}  d\vec{p}_{\bar{q}} \; \left[
3 \gamma \delta(\vec{p}_{q} + \vec{p}_{\bar{q}}) \sum_{m}
(110|m,-m) {\cal Y}_{1}^{m}(\vec{p}_{q} -\vec{p}_{\bar{q}})
{\cal Z}_{i,j}^{-m} \right] \cdot \nonumber \\ 
& & \;\;\;\;\; b^{\smdag}_{i}(\vec{p}_{q})
d^{\smdag}_{j}(\vec{p}_{\bar{q}})
\end{eqnarray}	

\noindent
where $b_{i}^{\smdag}$ ($d_{j}^{\smdag}$) are the $i$-quark
($j$-antiquark) creation operators, $\vec{p}_{q}$
($\vec{p}_{\bar{q}}$) is the three-momentum of the
quark (antiquark) of the pair, ${\cal Y}_{1}^{m}$ is the solid
harmonic polynomial, ${\cal Z}$ is the color--spin--isospin wave
function of the pair, (110$|$m,-m) is the SU(2) Clebsch--Gordan
coefficient and $\gamma$ is the coupling constant at the vacuum--$q\bar{q}$ 
vertex. Following our criterium given above, $\gamma$ is extracted from
$f_{NN\pi} (0)$.

	Then the transition matrix element reads: 

\begin{equation}
\label{3p0matrix}
\langle B' \pi  | T | B \rangle = - 3 \gamma \sum_{m} (1 1 0| m, -m)
I_{m}
\end{equation}  

\noindent 
where

\begin{eqnarray}
I_{m} & = & \delta(\vec{P}_{B} - \vec{P}_{B'} - \vec{k}) \int
d\vec{p}_{\xi_{1}} \; d\vec{p}_{\xi_{2}}  \nonumber \\
& & {\cal Y}_{1}^{m}
\left[ -\frac{4}{3} \vec{k} - \sqrt{\frac{2}{3}} \left(
\vec{p}_{\xi_{2}} + (\vec{p}_{\xi_{2}} + \sqrt{\frac{2}{3}} \vec{k}) \right) + 
\frac{2}{3} \vec{P}_{B} \right] \Phi_{\mbox {\footnotesize Pair}}^{-m} \cdot
\nonumber \\ 
& & \left[ \Psi_{B'}(\vec{p}_{\xi_{1}}, \vec{p}_{\xi_{2}} +
\sqrt{\frac{2}{3}} \vec{k})  \Phi_{B'} \right]^{*}_{J'\lambda} 
\left[ \Psi_{\pi}(- \sqrt{\frac{2}{3}} \vec{p}_{\xi_{1}} +
\frac{\vec{P}_{B}}{3} - \frac{\vec{k}}{2})  \Phi_{\pi} \right]^{*}  
\cdot \nonumber \\
&  & \left[ \Psi_{B}(\vec{p}_{\xi_{1}}, \vec{p}_{\xi_{2}})
\Phi_{B} \right]_{J\lambda}
\end{eqnarray}

\noindent 
where $\Phi_{\mbox {\footnotesize Pair}}^{-m}$ is the spin-isospin
wave function of the pair and $\Psi_{\pi}$ ($\Phi_{\pi}$) refers to
the momentum space (spin-isospin) wave function of the pion. We choose
for $\Psi_{\pi}$ a gaussian form 
with the parameter $R_{A}^{2} = 8$ GeV$^{-2}$ 
in order to reproduce the root mean square radius of the pion.

	 The decay widths, calculated elsewhere \cite{CANO95}, are compiled
in tables 3a and 3b. A comparison to the second column of tables 2a and
2b shows clearly the relevance of the pion structure.
However, the predicted values differ very much 
from the experimental data.
It is very illuminating to take the point-like pion limit of
the transition matrix element $<B'\pi|T|B>$. In ref. \cite{CANO95} this was
done in configuration space and compared to the $(p/m)$ order expression
suggesting a way to introduce relativistic--like
corrections in the $^{3}P_{0}$ scheme. The results obtained in this
manner (we shall call it modified $^{3}P_{0}$ model (M$^{3}P_{0}$))
shown in tables 3a and 3b, represent a rather amazing improvement of the
predictions.
Here we repeat the procedure in momentum space in order to get a
'more convergent' $(p/E)$ version of the modified $^{3}P_{0}$ model. 

	In the point-like limit we get:

\begin{eqnarray}
\langle B ' \pi | T | B \rangle & \rightarrow & \frac{1}{(2 \pi)^{3/2}}
(- \gamma 3 \sqrt{3} \pi)
\int d\vec{p}_{\xi_{1}} \;  d\vec{p}_{\xi_{2}} \nonumber \\ 
& & \left[ \frac{4}{3} \vec{k} 
+ \sqrt{\frac{2}{3}} \left[ \vec{p}_{\xi_{2}} +  
 (\vec{p}_{\xi_{2}} + \sqrt{\frac{2}{3}} \vec{k}) \right] \right]
\vec{\Phi}_{\mbox {\footnotesize Pair}} \cdot \nonumber \\
& & \left[ \Psi_{B'}(\vec{p}_{\xi_{1}},\vec{p}_{\xi_{2}}+
\sqrt{\frac{2}{3}} \vec{k})  \Phi_{B'} 
\right]_{J'\lambda}^{*} \Phi_{M}^{*}
\left[ \Psi_{B}(\vec{p}_{\xi_{1}},\vec{p}_{\xi_{2}}) \Phi_{B} 
\right]_{J\lambda} 
\end{eqnarray}

	This expression reproduces (\ref{amplitudgordape}) under the
formal substitutions:

\newcounter{Item}
\refstepcounter{equation}
\setcounter{Item}{\value{equation}}
\setcounter{equation}{0}
\renewcommand{\theequation}{\arabic{Item}.\alph{equation}}
\label{conditions}

\begin{eqnarray}
\label{condition1} 
3 \sqrt{3} \frac{\pi}{4} \gamma & \rightarrow &
\frac{3 i f_{qq\pi}}{(2 \omega_{\pi})^{1/2} m_{\pi}}\\
\frac{4}{3} & \rightarrow &  
\left( 1 + \frac{\omega_{\pi}}{6 E'} \right) \label{condition2} \\
\vec{p}_{\xi_{2}} +  
 (\vec{p}_{\xi_{2}} + \sqrt{\frac{2}{3}} \vec{k}) & \rightarrow & 
\frac{\omega_{\pi}}{2} \left[
\frac{\vec{p}_{\xi_{2}}}{E} +  
\frac{(\vec{p}_{\xi_{2}} + \sqrt{\frac{2}{3}} \vec{k})}{E'} 
\right] \label{condition3} 
\end{eqnarray}
\setcounter{equation}{\value{Item}}
\renewcommand{\theequation}{\arabic{equation}}

\noindent
Eq. (\ref{condition1}) points out an energy dependence (through the
factor $\omega_{\pi}^{1/2}$) of the coupling constant $\gamma$ that
can be associated to the boost from the
rest frame of the pion to the decaying baryon rest frame. Moreover the
expressions on the right hand side of (\ref{condition2}) and
(\ref{condition3}) reduce to the left hand side in the very
non--relativistic limit $E \sim E' \sim m$, $(\omega_{\pi})_{^{3}P_{0}}
\sim 2 m$. 

	Hence we can go the other way around and use the right hand side
of eq. (\ref{conditions}) as the relativistic expressions of the
left hand side, to modify the $^{3}P_{0}$ effective transition
operator. Technically the substitutions translate in the modification of
the argument of the spherical harmonic plus an energy dependent factor:

\begin{equation}
\begin{array}{rcl}
 &   {\cal Y}_{1}^{m} \left[ -\frac{4}{3} \vec{k} - \sqrt{\frac{2}{3}} \left(
\vec{p}_{\xi_{2}} + (\vec{p}_{\xi_{2}} + \sqrt{\frac{2}{3}} \vec{k})
\right) \right]  & \\ 
\rightarrow  &  \sqrt{\frac{m_{\pi}}{\omega_{\pi}}} 
{\cal Y}_{1}^{m} \left[ 
- \vec{k} \left( 1 + \frac{\omega_{\pi}}{6 E'}\right) 
 + \frac{\omega_{\pi}}{2}
\left( - \sqrt{\frac{2}{3}} \right) 
\left( \frac{\vec{p}_{\xi_{2}}}{E} +  
\frac{\vec{p}_{\xi_{2}} + \sqrt{\frac{2}{3}} \vec{k}}{E'} \right) \right]
\end{array} 
\end{equation}

\noindent
Thus, the transition operator evidences symmetry properties expected from 
the elementary coupling of the derivative of the pion field to a
current involving baryons.
This  'relativized' $^{3}P_{0}$ model that we shall call
R$^{3}P_{0}$ from now on confirm the good predictions the M$^{3}P_{0}$
version gave (the $NN\pi$ form factor is very similar
to the one obtained with the usual $^{3}P_{0}$ model).
 It is worthwhile to mention the
excellent fit we get for the radial decays $N(1440) \rightarrow N\pi$,
$\Delta(1600) \rightarrow \Delta\pi$, due entirely to the incorporation
of the pion structure not having any need to consider any exotic
nature for them. To this respect, it should be mentioned that the 
$(p/E)^2$ and higher order corrections to them, although they were not small
in the elementary emission model, are not very relevant compared to those
arising from the pion structure. We get a good fit (exceptions are the
$N(1535) \rightarrow \Delta \pi$ 
and $N(1650) \rightarrow \Delta \pi$) even
for some non-justified first-order $(p/E)$ treated cases maybe 
indicating the reabsortion of
some relativistic effects through the only parameter.
 
	The $N(1535) \rightarrow N\pi$ and 
$N(1650) \rightarrow N \pi$ decay widths are very badly described. 
For this several reasons can be pointed out. 
Higher order corrections are very significant  as can be easily 
inferred from the tables.
Besides, for these decays
one should be aware of the presence of threshold effects 
\cite{BERNARD} (the coupling for the 
$N\eta$ channel is important). Additionally the lack of a tensor
interaction at our model prevents mixing effects that could be relevant.
To this respect we have evaluated the mixing that a tensor potential
with the OGE parameter we have used would introduce. Though the mixing
angle ($\vartheta \approx -40^{\mbox o}$ 
is similar to the obtained with other
models it does not translate in ours in a general improvement of all
the $N(1535)$ and $N(1650)$ decays.

	In all cases, the effect of the emitted baryon width 
\cite{CAPSTICK94} and the
final state interactions which have been neglected might play some
role as well. Furthermore an extended pion-quark vertex could be
considered through a form factor multiplying $f_{qq\pi}$ (or
$\gamma$). 
An estimation of its importance for a standard dipole type
form factor \cite{ITONAGA} gives corrections to the widths 
of a 20 \% at most. It is also
worthwhile to realize the better predictions on the average that
three--quark potential provide. They support the results for the
baryonic spectrum and may indicate the convenience of the presence of
high momentum components in the baryon wave function at low energies,
though no strong conclusion should be derived from our restricted
calculation.

	Other treatments of strong decays can be found in the
literature within a $^3P_0$ scheme \cite{STANCU88} getting a reasonable
overall fit to a much wider range of data than considered here.
However the adopted philosophy is quite a different one; we put the
emphasis in the  $(p/E)$ expansion that allows the 
study of better relativistically controlled
processes and show that  a precise simultaneous description of the
spectrum and the decays is feasible in some cases.

\section{Summary.}

	We have performed a detailed analysis of strong pionic 
decay widths from a non--relativistic scheme, first following a $(p/m)$
expansion approach, second  using for the energy its relativistic
expression ($E=\sqrt{m^2 + p^2}$) instead of the $E \approx m$
non-relativistic limit. A clear improvement of the convergence is
obtained and a justification of a $(p/E)$ treatment is obtained for
some cases. The strong remaining discrepancies between our theoretical
predictions and experiment has led us to analyse through a $^{3}P_{0}$
pair creation model the influence of the $q\bar{q}$ content of the
pion in the decay width and the need of
introducing relativistic--like corrections into the effective
transition operator. These  relativistic corrections 
have basically a kinematical origin
An improvement of the $(p/m)$ results for the examined cases
is achieved in good  agreement with experimental data even for
some non--rapidly convergent processes. 
Let us mention in particular that the sizeable transitions between the
$N$ and $\Delta$ ground states and their radial excitations originate 
from the pion structure and not from the $(p/m)^2$ terms in the operator
as one may naively expect from the fact that the orthogonality of the 
states is of no relevance in this case. The validity of the argument 
is restored when the full series of higher order terms is considered.
Being
conscious of possible minor improvements that could be considered, as
a general conclusion we might say that in order to do better for the
pionic decay widths a more complete treatment of relativistic effects
and the $q\bar{q}$ structure altogether in a consistent scheme seems
to be unavoidable

	This work has been partially supported by DGES under grants
PB95-1096, PB94-0080 and by EC-TMR network
HaPHEEP under contract ERBTMRX CT96-0008. F.C. acknowledges the
Ministerio de Educaci\'on y Ciencia for a FPI fellowship.


\newpage 


\begin{center}
{\bf Table captions}
\end{center}

\begin{description}

\item{\bf Table 1}  Fitted values of the parameters of the
two--body and two+three body potentials. 

\item{\bf Table 2a} Pionic decay widths and the $f_{qq\pi}$ coupling constant
obtained with the elementary
emission model (EEM) for the two--body quark potential model. 
Experimental data from \protect{\cite{PDG94}}.

\item{\bf Table 2b} Pionic decay widths and the $f_{qq\pi}$ coupling constant
obtained with the elementary
emission model (EEM) for the two+three body quark potential model.

\item{\bf Table 3a} Pionic decay widths and the $\gamma$ constant
obtained with several versions
of the $^3P_{0}$ model for the two--body quark potential model. 

\item{\bf Table 3b} Pionic decay widths and the $\gamma$ constant
obtained with several versions
of the $^3P_{0}$ model for the two+three body quark potential model. 

\end{description}

\newpage


\begin{table}[p]
\begin{center}
\begin{tabular}{|c|c|c|c|}
\hline \hline 
\multicolumn{2}{|c|}{\rule{0pt}{3.5ex}}
& $V^{(2)}$ & $V^{(2)}+V^{(3)}$ \\[2ex]
\hline
\multicolumn{2}{|c|}{\rule{0pt}{3.5ex}$m_{u}=m_{d}$ (GeV)} 
& 0.337 & 0.355  \\[2.5ex]
\hline
\rule{0pt}{3.5ex}
$V^{\mbox {\scriptsize (COUL)}}$ & $\kappa$ (GeV fm) 
& 0.1027 & 0.289  \\[2.5ex]
\hline
$V^{(\vec{\sigma} \vec{\sigma})}$ & 
\rule{0pt}{4.5ex} 
\begin{tabular}{l} $\kappa_{\sigma}$ (GeV fm) \\ $r_{0}$ (fm) 
\end{tabular} &
\begin{tabular}{c} 0.1027 \\ 0.4545 
\end{tabular} &
\begin{tabular}{c} 0.049 \\ 0.40
\end{tabular} \\[2.5ex]
\hline
\rule{0pt}{3.5ex}
$V^{\mbox {\scriptsize (CONF)}}$ & $a^{2}$ (GeV$^{-1}$ fm) 
& 1.063 & 4.570 \\[2.5ex]
\hline
\rule{0pt}{4.5ex}
$V^{(3)}$ & 
\begin{tabular}{l} $V_{0}$ (GeV$^{-2}$ fm$^{-6}$) \\ $m_{0}$ (GeV) 
\end{tabular} &
\begin{tabular}{c} -- \\ --
\end{tabular} &
\begin{tabular}{c} -61.63 \\ 0.25
\end{tabular} \\[2.5ex]
\hline\hline
\end{tabular}
\vskip 2cm
{\bf Table 1}
\end{center}
\end{table}

\newpage


\begin{table}
\begin{center}
{\tabcolsep 3pt
\begin{tabular}{@{}lcccccc}
\hline \hline
\multicolumn{1}{c}{\rule{0pt}{5.5ex} {\large $V^{(2)}$}} & 
\begin{tabular}{c} EEM  \\ $(p/m)$ 
\end{tabular} & 
\begin{tabular}{c} EEM \\ $(p/m)^{2}$ 
\end{tabular}  & 
\begin{tabular}{c} EEM \\ $(p/E)$ 
\end{tabular} & 
\begin{tabular}{c} EEM \\ $(p/E)^{2}$
\end{tabular} &
\begin{tabular}{c} EEM \\  {\scriptsize All $(p/E)$ orders}  
\end{tabular} &    
\begin{tabular}{c} $\Gamma_{\mbox {\scriptsize Exp}}$ \\ (MeV)
\end{tabular}  \\ [3ex]
\hline \hline
\rule{0pt}{4.5ex}
$\Delta(1232) \rightarrow N \pi$ & 
79.6 & 25.1 & 74.3 & 75.0 & 75.8 & 115--125 \\ [2.ex]
\rule{0pt}{2.5ex}
$N(1440) \rightarrow N \pi$ & 
 3.4 & 177 & 0.007 & 4.91 & 13.1 & 210--245 \\ [2.ex]
\rule{0pt}{2.5ex}
$N(1440) \rightarrow \Delta \pi$ & 
7.1 & 13.3  & 5.4 & 7.82 & 9.30 & 70--105 \\ [2.ex]
\rule{0pt}{2.5ex}
$\Delta(1600) \rightarrow N \pi$ & 
20.1 & 94.1 & 31.0 & 15.4 & 7.98 & 35--88 \\ [2.ex]
\rule{0pt}{2.5ex}
$\Delta(1600) \rightarrow \Delta \pi$ & 
2.85 & 56.4 & 0.72 & 5.08 & 9.30 &140--245 \\ [2.ex]
\rule{0pt}{2.5ex}
$N(1520) \rightarrow N \pi$ & 
61.8 & 21.5 & 62.3 & 60.6 & 57.7 & 60--72  \\ [2.ex]
\rule{0pt}{2.5ex}
$N(1520) \rightarrow \Delta \pi$ & 
78.0 & 95.3 & 24.3 &  39.2 & 45.1 & 18--30 \\ [2.ex]
\rule{0pt}{2.5ex}
$N(1535) \rightarrow N \pi$ & 
240 & 494 & 24.8 & 76.5 & 101 & 53--83 \\ [2.ex]
\rule{0pt}{2.5ex}
$N(1535) \rightarrow \Delta \pi$ & 
9.7 & 4.25  & 9.93 & 10.4 & 10.3 & $< $1.5 \\ [2.ex]
\rule{0pt}{2.5ex}
$N(1650) \rightarrow N \pi$ & 
47.9 & 135  & 2.49 & 14.5 & 20.9 & 90--120 \\ [2.ex]
\rule{0pt}{2.5ex}
$N(1650) \rightarrow \Delta \pi$ & 
12.4 & 5.49  & 12.6 & 13.0 & 12.8 & 4--11 \\ [2.ex]
\rule{0pt}{2.5ex}
$N(1700) \rightarrow N \pi$ & 
4.07 & 1.22  & 3.65 & 3.43 & 3.24 & 5--15 \\ [2.ex]
\rule{0pt}{2.5ex}
$N(1700) \rightarrow \Delta \pi$ & 
383 & 612 & 112 & 190 & 222 & 81--393 \\ [2.ex]
\hline
$f_{qq\pi}$ & 0.602 & 0.602 & 0.619 & 0.743 & 0.782 &  \\
\hline\hline	
\end{tabular}}
\vskip 2cm
{\bf Table 2a}
\end{center}
\label{table1}
\end{table}

\newpage


\begin{table}
\begin{center}
{\tabcolsep 3pt
\begin{tabular}{@{}lcccccc}
\hline \hline
\multicolumn{1}{c}{\rule{0pt}{5.5ex} {\large $V^{(2)}+V^{(3)}$}} & 
\begin{tabular}{c} EEM  \\ $(p/m)$ 
\end{tabular} & 
\begin{tabular}{c} EEM \\ $(p/m)^{2}$ 
\end{tabular}  & 
\begin{tabular}{c} EEM \\ $(p/E)$ 
\end{tabular} & 
\begin{tabular}{c} EEM \\ $(p/E)^{2}$
\end{tabular} & 
\begin{tabular}{c} EEM \\  {\scriptsize All $(p/E)$ orders} 
\end{tabular} & 
\begin{tabular}{c} $\Gamma_{\mbox {\scriptsize Exp}}$ \\ (MeV)
\end{tabular}  \\ [3ex]
\hline \hline
\rule{0pt}{4.5ex}
$\Delta(1232) \rightarrow N \pi$ & 
72.1 & 4.18 & 67.4 & 71.7 & 75.7 & 115--125 \\ [2.ex]
\rule{0pt}{2.5ex}
$N(1440) \rightarrow N \pi$ & 
0.17 & 690 & 0.04 & 4.01 & 15.3 & 210--245 \\ [2.ex]
\rule{0pt}{2.5ex}
$N(1440) \rightarrow \Delta \pi$ & 
17.6 & 43.1  & 16.7 & 24.5 & 30.9 & 70--105 \\ [2.ex]
\rule{0pt}{2.5ex}
$\Delta(1600) \rightarrow N \pi$ & 
94.1 & 226 & 84.9 & 64.1 & 50.4  & 35--88 \\ [2.ex]
\rule{0pt}{2.5ex}
$\Delta(1600) \rightarrow \Delta \pi$ & 
0.10 & 70.4 & 0.006 & 2.08 & 5.89 & 140--245 \\ [2.ex]
\rule{0pt}{2.5ex}
$N(1520) \rightarrow N \pi$ & 
22.3 & 15.3 & 25.9 & 29.3 & 31.5 & 60--72  \\ [2.ex]
\rule{0pt}{2.5ex}
$N(1520) \rightarrow \Delta \pi$ & 
56.1 & 67.4 & 20.4 &  36.2 & 44.8 & 18--30 \\ [2.ex]
\rule{0pt}{2.5ex}
$N(1535) \rightarrow N \pi$ & 
149 & 259 & 20.7 & 60.7 & 82.8 & 53--83 \\ [2.ex]
\rule{0pt}{2.5ex}
$N(1535) \rightarrow \Delta \pi$ & 
8.3 & 5.78  & 9.25 & 11.3 & 12.6 & $< $1.5 \\ [2.ex]
\rule{0pt}{2.5ex}
$N(1650) \rightarrow N \pi$ & 
24.9 & 60.3  & 2.54 & 11.0 & 15.8 & 90--120 \\ [2.ex]
\rule{0pt}{2.5ex}
$N(1650) \rightarrow \Delta \pi$ & 
9.95 & 6.35  & 10.4 & 12.3 & 13.6 & 4--11 \\ [2.ex]
\rule{0pt}{2.5ex}
$N(1700) \rightarrow N \pi$ & 
1.43 & 0.75  & 1.29 & 1.41 & 1.52 & 5--15 \\ [2.ex]
\rule{0pt}{2.5ex}
$N(1700) \rightarrow \Delta \pi$ & 
220 & 347  & 79.7 & 148 & 185 & 81-393 \\ [2.ex]
\hline
$f_{qq\pi}$ & 0.604 & 0.604 & 0.624 & 0.780 & 0.852 &  \\
\hline\hline	
\end{tabular}}
\vskip 2cm
{\bf Table 2b}
\end{center}
\label{table2}
\end{table}


\begin{table}
\begin{center}
{\tabcolsep 10pt
\begin{tabular}{@{}lcccc}
\hline \hline
\multicolumn{1}{c}{\rule{0pt}{5.5ex} {\large $V^{(2)}$}}
& $^{3}P_{0}$ & M$^{3}P_{0}$ & R$^{3}P_{0}$ 
& $\Gamma_{\mbox {\scriptsize Exp}}$ (MeV)  \\[2ex]
\hline \hline
\rule{0pt}{4.5ex}
$\Delta(1232) \rightarrow N \pi$ & 
167 & 88.6 & 83.9 & 115--125 \\ [2.ex]
\rule{0pt}{2.5ex}
$N(1440) \rightarrow N \pi$ & 
452 & 114 & 73.5 & 210--245 \\ [2.ex]
\rule{0pt}{2.5ex}
$N(1440) \rightarrow \Delta \pi$ & 
66.5 & 27.6 & 20.7 & 70--105 \\ [2.ex]
\rule{0pt}{2.5ex}
$\Delta(1600) \rightarrow N \pi$ & 
19.8 & 2.1 & 0.27 & 35--88 \\ [2.ex]
\rule{0pt}{2.5ex}
$\Delta(1600) \rightarrow \Delta \pi$ & 
255  & 62.0 & 41.9 & 140--245 \\ [2.ex]
\rule{0pt}{2.5ex}
$N(1520) \rightarrow N \pi$ & 
268 & 95.1 & 92.0 & 60--72  \\ [2.ex]
\rule{0pt}{2.5ex}
$N(1520) \rightarrow \Delta \pi$ & 
532 & 45.9 & 17.1 & 18--30 \\ [2.ex]
\rule{0pt}{2.5ex}
$N(1535) \rightarrow N \pi$ & 
429 & 49.2 & 0.18 & 53--83 \\ [2.ex]
\rule{0pt}{2.5ex}
$N(1535) \rightarrow \Delta \pi$ & 
28.1 &  15.3 & 15.7 & $< $1.5 \\ [2.ex]
\rule{0pt}{2.5ex}
$N(1650) \rightarrow N \pi$ & 
49.1  & 6.14  & 1.13 & 90--120 \\ [2.ex]
\rule{0pt}{2.5ex}
$N(1650) \rightarrow \Delta \pi$ & 
49.3 & 20.9  & 20.9 & 4--11 \\ [2.ex]
\rule{0pt}{2.5ex}
$N(1700) \rightarrow N \pi$ & 
11.5 & 3.49  & 3.15 & 5--15 \\ [2.ex]
\rule{0pt}{2.5ex}
$N(1700) \rightarrow \Delta \pi$ & 
1643 & 228 & 114 & 81-393 \\ [2.ex]
\hline
$-i \gamma$ & 7.02 & 8.20 & 8.29 & \\
\hline\hline	
\end{tabular}}
\vskip 2cm
{\bf Table 3a}
\end{center}
\label{table3}
\end{table}


\begin{table}
\begin{center}
{\tabcolsep 10pt
\begin{tabular}{@{}lcccc}
\hline \hline
\multicolumn{1}{c}{\rule{0pt}{5.5ex} {\large $V^{(2)}+V^{(3)}$}}
& $^{3}P_{0}$ & M$^{3}P_{0}$ & R$^{3}P_{0}$ 
& $\Gamma_{\mbox {\scriptsize Exp}}$ (MeV)  \\[2ex]
\hline \hline
\rule{0pt}{4.5ex}
$\Delta(1232) \rightarrow N \pi$ & 
210 & 112 & 106 & 115--125 \\ [2.ex]
\rule{0pt}{2.5ex}
$N(1440) \rightarrow N \pi$ & 
 1076 & 307 & 236 & 210--245 \\ [2.ex]
\rule{0pt}{2.5ex}
$N(1440) \rightarrow \Delta \pi$ & 
228 & 116 & 106 & 70--105 \\ [2.ex]
\rule{0pt}{2.5ex}
$\Delta(1600) \rightarrow N \pi$ & 
0.53 & 2.5 & 8.82 & 35--88 \\ [2.ex]
\rule{0pt}{2.5ex}
$\Delta(1600) \rightarrow \Delta \pi$ & 
498 & 121 & 93.5 & 140--245 \\ [2.ex]
\rule{0pt}{2.5ex}
$N(1520) \rightarrow N \pi$ & 
319 & 105 & 100 & 60--72  \\ [2.ex]
\rule{0pt}{2.5ex}
$N(1520) \rightarrow \Delta \pi$ & 
999 & 75.2 & 34.4 & 18--30 \\ [2.ex]
\rule{0pt}{2.5ex}
$N(1535) \rightarrow N \pi$ & 
464 & 44.1 & 0.52 & 53--83 \\ [2.ex]
\rule{0pt}{2.5ex}
$N(1535) \rightarrow \Delta \pi$ & 
74.0 &  36.3 & 37.7 & $< $1.5 \\ [2.ex]
\rule{0pt}{2.5ex}
$N(1650) \rightarrow N \pi$ & 
44.2  & 4.01  & 0.77 & 90--120 \\ [2.ex]
\rule{0pt}{2.5ex}
$N(1650) \rightarrow \Delta \pi$ & 
109 & 45.7 & 43.1 & 4--11 \\ [2.ex]
\rule{0pt}{2.5ex}
$N(1700) \rightarrow N \pi$ & 
11.2 & 3.25  & 2.83 & 5--15 \\ [2.ex]
\rule{0pt}{2.5ex}
$N(1700) \rightarrow \Delta \pi$ & 
2417 & 323 & 217& 81-393 \\ [2.ex]
\hline
$-i \gamma$ & 9.78 & 11.4 & 11.6 & \\
\hline\hline	
\end{tabular}}
\vskip 2cm
{\bf Table 3b}
\end{center}
\label{table4}
 \end{table}


\begin{thebibliography}{99}

\bibitem{BARYON95} Proc. of Baryons' 95 Conference, Santa Fe, 1995,
                    ed. B.F. Gibson et al., (World Scientific, 1996).

\bibitem{CAPSTICK93} S. Capstick and W. Roberts, Phys Rev D47 (1993) 1994.


\bibitem{CANO95} F. Cano et al.,
		Nucl. Phys A603 (1996) 257.  

\bibitem{DESPLANQUES92} B. Desplanques et al., Z. Phys. A343 
		(1992) 331.

\bibitem{GIGNOUX85} B. Silvestre--Brac and C. Gignoux, Phys. Rev. D37
			(1985) 74.  \newline
			R.K. Bhaduri et al., Nuovo Cim. A65 (1981) 376.

\bibitem{PDG94} Particle Data Group, Phys. Rev. D50 (1994) 1173.
 
\bibitem{CANO94} F. Cano, Master Thesis, Univ. of Valencia (1994).

\bibitem{LEYAOUANC73} A. Le Yaouanc et al.,
			 Phys. Rev. D8 (1973) 2223;
			 D11 (1975) 1272.

\bibitem{ROBERTS92} W. Roberts and B. Silvestre--Brac,
			Few Body Sys. 11 (1992) 171.

\bibitem{BERNARD} B. Silvestre--Brac and C. Gignoux, Phys. Rev. D43
                        (1991) 3699.

\bibitem{CAPSTICK94} S. Capstick and W. Roberts,
			 Phys. Rev. D49 (1994) 4570.

\bibitem{ITONAGA} K. Itonaga et al., Nucl. Phys. A609 (1996) 422.

\bibitem{STANCU88} Fl. Stancu and P. Stassart, Phys. Rev. D38 (1988) 233.
\end{thebibliography}
\end{document}